\documentclass[a4paper,fleqn,usenatbib]{mnras}

\usepackage{newtxtext,newtxmath}

\usepackage[T1]{fontenc}
\usepackage{ae,aecompl}
\usepackage{graphicx}	% Including figure files
\usepackage{amsmath}	% Advanced maths commands
\usepackage{amssymb}	% Extra maths symbols
\usepackage{soul}
\usepackage{ulem}
\usepackage{subfloat}
\title[Torque reversals of neutron stars]{ON THE TORQUE REVERSALS OF ACCRETING NEUTRON STARS}

\author[\"{U}. Ertan]{
\"{U}nal  Ertan \thanks{E-mail: unal@sabaciuniv.edu}
\\
Sabanc\i\ University, 34956, Orhanl\i\, Tuzla, \.Istanbul,
Turkey}
\date{Accepted XXX. Received YYY; in original form ZZZ}

\pubyear{2020}

\begin{document} 
\label{firstpage}
\pagerange{\pageref{firstpage}--\pageref{lastpage}}
\maketitle

\def\be{\begin{equation}}
\def\ee{\end{equation}}
\def\ba{\begin{eqnarray}}
\def\ea{\end{eqnarray}}
\def\m{\mathrm}
\def\d{\partial}
\def\R{\right}
\def\L{\left}
\def\a{\alpha}
\def\acold{\alpha_\mathrm{cold}}
\def\ahot{\alpha_\mathrm{hot}}
\def\Mdotstar{\dot{M}_\ast}
\def\Omegastar{\Omega_\ast}
\def\Omegadot{\dot{\Omega}}
\def\OmegaK{\Omega_{\mathrm{K}}}
\def\Mdotin{\dot{M}_{\mathrm{in}}}
\def\Mdots{\dot{M}_{\mathrm{s}}}

\def\Mdotcrit{\dot{M}_{\mathrm{crit}}}
\def\Mdotout{\dot{M}_{\mathrm{out}}}

\def\Mdot{\dot{M}}
\def\Edot{\dot{E}}
\def\Pdot{\dot{P}}
\def\nudot{\dot{\nu}}
\def\Msun{M_{\odot}}
\def\Lin{L_{\mathrm{in}}}
\def\Lcool{L_{\mathrm{cool}}}
\def\Mdotstar{\dot{M}_\ast}
\def\Rstar{R_\ast}
\def\rstar{r_\ast}

\def\Lstar{L_\ast}

\def\Rin{R_{\mathrm{in}}}

\def\rin{r_{\mathrm{in}}}
\def\rlc{r_{\mathrm{LC}}}
\def\rout{r_{\mathrm{out}}}
\def\rco{r_{\mathrm{co}}}
\def\re{r_{\mathrm{e}}}
\def\Ldisk{L_{\mathrm{disk}}}
\def\Lx{L_{\mathrm{x}}}
\def\Ld{L_{\mathrm{d}}}
\def\Lxpulsed{L_{\mathrm{x,pulsed}}}
\def\Md{M_{\mathrm{d}}} 
\def\NH{N_{\mathrm{H}}}
\def\dEb{\delta E_{\mathrm{burst}}}
\def\dEx{\delta E_{\mathrm{x}}}
\def\Bstar{B_\ast}\def\uff{\upsilon_{\mathrm{ff}}}
\def\Bb{\beta_{\mathrm{b}}}
\def\Be{B_{\mathrm{e}}}
\def\Bp{B_{\mathrm{p}}}
\def\Bz{B_{\mathrm{z}}}
\def\Bfi{B_{\mathrm{|phi}}}
\def\BA{B_{\mathrm{A}}}
\def\tint{t_{\mathrm{int}}}
\def\tdiff{t_{\mathrm{diff}}}
\def\r_m{r_{\mathrm{m}}}
\def\rA{r_{\mathrm{A}}}
\def\RA{R_{\mathrm{A}}}
\def\BA{B_{\mathrm{A}}}
\def\rS{r_{\mathrm{S}}}
\def\rp{r_{\mathrm{p}}}
\def\rxi{r_{\mathrm{\xi}}}
\def\Rxi{R_{\mathrm{\xi}}}
\def\reta{r_{\mathrm{\eta}}}
\def\Reta{R_{\mathrm{\eta}}}
\def\Mdoteta{\dot{M}_{\mathrm{\eta}}}
\def\Mdotxi{\dot{M}_{\mathrm{\xi}}}
\def\Tp{T_{\mathrm{p}}}
\def\dMin{\delta M_{\mathrm{in}}}
\def\Rc{\R_{\mathrm{c}}}
\def\Teff{T_{\mathrm{eff}}}
\def\uff{\upsilon_{\mathrm{ff}}}
\def\Tirr{T_{\mathrm{irr}}}
\def\Firr{F_{\mathrm{irr}}}
\def\Tcrit{T_{\mathrm{crit}}}
\def\P0min{P_{0,{\mathrm{min}}}}
\def\Av{A_{\mathrm{V}}}
\def\ah{\alpha_{\mathrm{hot}}}
\def\ac{\alpha_{\mathrm{cold}}}
\def\tc{\tau_{\mathrm{c}}}
\def\p{\propto}
\def\m{\mathrm}
\def\fast{\omega_{\ast}}
\def\Uff{\upsilon_{\mathrm{ff}}}
\def\Ufi{\upsilon_{\fi}}
\def\Ur{\upsilon_{\mathrm{r}}}
\def\UK{\upsilon_{\mathrm{K}}}
\def\Uesc{\upsilon_{\mathrm{esc}}}
\def\Uout{\upsilon_{\mathrm{out}}}
\def\Uphi{\upsilon_{\phi}}
\def\Udiff{\upsilon_{\mathrm{diff}}}
\def\Ure{\upsilon_{\mathrm{r,e}}}
\def\U{\upsilon}
\def\UB{\upsilon_{\mathrm{B}}}
\def\tauB{\tau_{\mathrm{B}}}
\def\hA{h_{\mathrm{A}}}
\def\he{h_{\mathrm{e}}}
\def\cs{c_{\mathrm{s}}}
\def\cse{c_{\mathrm{s,e}}}
\def\hin{h_{\mathrm{in}}}
\def\rhop{\rho^{\prime}}
\def\rhod{\rho_\mathrm{d}}
\def\rhos{\rho_\mathrm{s}}
\def\rhodp{\rho_\mathrm{d}^{\prime}}
\def\rhoe{\rho_\mathrm{e}}
\def\rhoout{\rho_\mathrm{out}}
\def\Alfven{Alfv$\acute{\mathrm{e}}$n~}
\def\418{SGR 0418+5729}
\def\142{AXP 0142+61}
\def\Caliskan{\c{C}al{\i}\c{s}kan~}
\def\ql{\textquotedblleft}
\def\qr{\textquotedblright~}

\def\gpers{g s$^{-1}$}
\def\ergpers{erg s$^{-1}$}
\def\Hzpers{Hz s$^{-1}$}
\def\spers{s s$^{-1}$}
\def\rinmax{r_{\mathrm{in,max}}}
\def\Rinmax{R_{\mathrm{in,max}}}

\begin{abstract}

We have extended the analytical model proposed earlier to estimate the inner disk radius of accreting neutron stars in  the strong-propeller (SP) phase, and the conditions for the transitions between the strong and weak propeller (WP) phases (Ertan 2017, 2018) to the WP  (accretion with spin-down) and the spin-up (SU) phases, and the torque reversals during the WP/SU transitions.   The model can account for some basic observed properties of these systems that are not expected in conventional models: (1) accretion on to the star at low X-ray luminosities and the transitions to the SP phase (no accretion) at critical accretion rates much lower than the rate required for the spin-up/spin-down transition, (2) ongoing accretion throughout a large range of accretion rates while the source is spinning down (WP phase), and (3) transitions between the spin-up and spin-down phases with comparable  torque magnitudes,  without substantial changes in the mass-flow rate.  Our results indicate that the magnitudes of the torques on either side of the torque reversal  have a ratio similar for different systems independently of their spin periods, magnetic dipole moments  and accretion rates during the transitions. Estimated torque reversal properties in our model are in agreement with the observed torque reversals of 4U 1626--67.

\end{abstract}

\begin{keywords}
pulsars: individual (4U 1626--67) -- accretion -- accretion disks
\end{keywords}

%%%%%%%%%%%%%%%%%%%%%%%%%%%%%%%%%%%%%%%%%%%%%%%%%%

%%%%%%%%%%%%%%%%% BODY OF PAPER %%%%%%%%%%%%%%%%%%

\section{Introduction}

Neutron stars interacting with accretion disks have three different regimes depending on the mass inflow rate $\Mdotin$: (1) at the lowest $\Mdotin$ a strong-propeller regime in which no mass accretion on to the star takes place and the neutron star is spinning down, (2) at intermediate $\Mdotin$  a weak propeller regime (accretion with spin-down) in which most or part of the matter is accreted on to the star, and the star is spinning down, and (3) at the highest $\Mdotin$, a spin-up regime in which all of the matter flowing in from the outer disk is accreted on to star. The critical values of  $\Mdotin$  where the transitions between these different regimes take place and the corresponding locations of the inner disk radius $\rin$ depend on the magnetic dipole moment $\mu$ and the rotation rate $\Omegastar$ of the neutron star. Recently, Ertan (2017, 2018) showed that  $\rin$ does not track, and is much smaller than the conventional \Alfven radius, $\rA$, and has a weak dependence on $\Mdotin$  in the strong-propeller phase. In this model, $\rin$ tracks the co-rotation radius, $\rco$, where the closed field lines rotate with the Kepler speed, in the weak-propeller phase.

This understanding of the transitions between the strong and the weak propeller regimes is successful in explaining the properties and torque variation of transitional millisecond pulsars  (tMSPs) during  their transitions between the radio pulsar and the X-ray pulsar  states (Archibald et al. 2009, Papitto et al. 2013,  Bassa et al. 2014, Jaodand et al. 2016). In these systems, the accretion on to the star persists at X-ray luminosities much lower than the critical  spin-up/spin-down transition  level   (Papitto et al. 2015, Archibald et al. 2015) estimated  in the conventional models (Illarionov \& Sunyaev 1975, Ghosh \& Lamb 1979).   
Hereafter, we will denote the weak-propeller, strong-propeller and spin-up phases by "WP", "SP", and "SU" respectively.

While the properties of tMSPs constrain the models for the SP/WP transitions,  a few strongly magnetized   ($\mu > 10^{29}$ G cm$^3$) accreting pulsars in LMXBs with relatively long spin periods provide laboratories to study the transitions between the WP and SU phases, that is, torque reversals. Among these LMXBs,   4U 1626--67 (Chakrabarty et al. 1997a), Her X-1, (Deeter at al. 1989, Wilson et al. 1994), GX 1+4 (Chakrabarty et al. 1997b) showed torque reversals, transitions between the SU and WP phases. Some X-ray pulsars in high mass X-ray binaries (HMXBs) also show torque reversals (Bildsten et al. 1997, \.Inam et al. 2009). Nevertheless,  HMXBs are not very convenient to study the details of the disk-field interaction due to the effect of the wind from the companion. Among the LMXBs that show spin-up/spin-down transitions, GX 1+4 (Hinkle et al. 2006) is also thought to accrete from the wind of its companion possibly with transient disk formation (Camero-Arranz et al. 2010). Her X-1 has a persistent disk, but varying attenuation of the pulsar due to near edge-on view does not allow a detailed study of its X-ray luminosity and the torque relation (Petterson et al. 1991). 4U 1626--67 seems to be the best source to study the details of the disk-field interaction leading to observed  torque reversals, while similarities in the torque reversal properties of other LMXBs, and possibly HMXBs with disk-fed neutron stars could also give hints for the mechanism of these transitions. 
 
There are several common, striking properties of LMXBs showing torque reversals which are not addressed by the conventional models (see Bildsten et al. 1997 for a review): (1) the magnitudes of the torques before and after the torque reversal are similar, (2) the accretion on to the star goes on while the star is spinning down, (3) torque reversals do not require a significant change in the mass accretion rate.  In this work,  to explain these basic properties of neutron stars interacting with geometrically thin accretion disks, we extend the model developed earlier by Ertan (2017, 2018) to include the SP, WP and SU phases, and the transitions between these phases in a single picture. We will test our model results with the torque reversal properties of 4U 1626--67, and also give the results for illustrative model sources with different periods and magnetic dipole moments. 
In Section 2, we describe the details of the model.  In Section 3, we discuss our results with examples and  comparisons with the observations of 4U 1626--67. Our conclusions are summarized in Section 4.

\section{The Model}

 Conventional \Alfven radius, $\rA$ is calculated by equating the magnetic pressure of the dipole field of the neutron star to the ram pressure of matter accreting with spherical symmetry on to the neutron star, which gives  $\rA \simeq (GM)^{1/7} \mu^{4/7} \Mdot^{-2/7}$ (Davidson \& Ostriker 1973, Lamb et al. 1973) where G is the gravitational constant, $M$ and $\mu$ are the mass and magnetic dipole moment of the neutron star, and $\Mdot $ is the mass accretion rate.  
The conditions in the case of accretion from a geometrically thin disk are rather different from the spherical accretion. At a given radius $r$, the disk matter moves with Kepler speed, $\UK = r  \OmegaK$, with a mass density orders of magnitude greater than in the spherical accretion  at a given radius. Furthermore, the radial speed of matter in the disk is orders of magnitude smaller than in the spherical accretion. Despite these differences, the radius at which the magnetic and viscous stresses are balanced in the disk accretion is estimated  to be very close to $\rA$ within a factor of 2 (Ghosh \& Lamb 1979, Arons 1993, Ostriker \& Shu 1995, Wang 1998). Hereafter, we denote this radius by $\rxi = \xi \rA$, and in our calculations we set $\xi = 0.5$ as estimated by Ghosh \& Lamb (1979). In conventional models, the inner disk radius $\rin$ in a steady-state is usually assumed to be equal to  $\rxi$.  

The diffusion timescale of the field lines in the disk is comparable to the viscous timescale which is much longer than the interaction  timescale of the field lines and the inner disk, $\tint = |\OmegaK - \Omegastar|^{-1}$, (Fromang \& Stone 2009) where $\Omegastar$   is the angular velocity of the neutron star. The field lines  cannot slip through the disk. Theoretical studies and numerical simulations show that the field lines interacting with the inner disk in a narrow boundary inflate and open up within the interaction timescale    (Aly 1985, Lovelace et al. 1995, Hayashi et al. 1996, Miller \& Stone 1997, Uzdensky et al. 2002, Uzdensky 2004).  If the system is in the SP phase, the matter flowing in to the boundary at the innermost region of the disk is expelled from the system along the open field lines. The open lines reconnect  on the dynamical timescale, $\OmegaK^{-1}$, and continue to apply torque on the matter until they open up again (Lovelace et al. 1999, Ustyugova et al. 2006). The simulations show that the field lines outside the interaction boundary are decoupled from the disk. The magnetosphere inside the boundary is the region where the closed field lines and the plasma can rotate together at radius $r_{\mathrm{m}} \simeq \rin$.  These results indicate that a strong-propeller mechanism could be sustained if the field lines expel the matter from the inner boundary at the same rate as that of the mass-flow from the outer disk. Furthermore, for this strong-propeller phase to be steady, the matter should be accelerated to speeds greater than the escape speed,  $\Uesc$, within $\tint$.  Ertan (2017) showed that the maximum radius at which the strong-propeller condition is satisfied is much smaller than $\rA$, while  the accretion rates estimated for the WP/SP transition in this model seem to be in agreement with the transition properties of tMSPs (Ertan 2017, 2018). The  inner boundary in the spin-up phase is also continuously  evacuated, as in the case of strong-propeller, but now due to accretion on to the star. At $\rin$ the field lines decelerate and bring the matter into co-rotation within $\tint$ in the spin-up phase. In other words, the same equation could determine the inner disk radius in both the strong propeller phase (for $\rin > \rco$) and the spin-up phase (for $\rin < \rco$).  However,  this is not the whole story,  since the conditions in the spin-up (SU) phase are rather different,  in particular $\rA$ enters the picture (Sections 2.1 - 2.4).     

In the model, mass accretion on to the star is allowed (WP and SU phases) when $\rin \le \rco$,  and the system is in the  SP  phase (no accretion on to the star) when $\rin > \rco$ where $\rco = (GM / \Omegastar^{2})^{1/3} $  is the co-rotation radius at which $\OmegaK = \Omegastar$. For a steady strong-propeller mechanism the field should cut the inner disk at a radius greater than  $r_1 = 1.26 ~\rco$, because the field lines inside this radius cannot expel matter from the boundary with speeds greater than $\Uesc$. In the WP phase, $\rin = \rco$ and the mass coupling to the field lines at this radius flows on to the star along the field lines. 
Below, starting from the source properties in the SP phase (Section 2.1), we describe the variation of the inner disk radius $\rin$  with increasing disk mass-flow rate $\Mdotin$, leading to the transitions  into the WP phase at a certain $\Mdotin$ (Section 2.2), and  into the SU phase at a much higher accretion rate (Section 2.3).  The torque calculations are described in Section 2.4.

\subsection{Strong-propeller (SP) phase}

In this phase, all the mass flowing into the narrow inner boundary region with radial width $\delta r$ is efficiently thrown out of the system with a rate equal to the rate of mass flow from the outer disk. $\delta r$ is estimated to be  a few disk thickness, much smaller than the $\rin$ (Ertan 2018). This boundary region is the continuously evacuated innermost region of the disk. The field lines could interact with the inner disk in a larger region outside the inner boundary with radial thickness $\Delta r$  that is a small fraction of $\rin$     ($\delta r \ll \Delta r <  r$).  This is likely to result in a flow of matter expelled from the larger boundary back to the disk at larger radii (backflow) in addition to the outflow from the inner boundary  (see Fig. 1 in Ertan 2018). In this case, a steady state is reached with a pile-up outside $\rin$, with a net mass-flow rate  $\Mdotin$  into $\delta r$ at $\rin$ equal to the rate of outflow from this region. Continuous interaction of the field lines  with the pile-up exerts a spin-down torque on the star that is much greater than the torque associated with the angular momentum loss to the outflowing gas (Ertan 2017).   

Through simple analytical calculations, Ertan (2017) showed that the maximum inner disk radius,  $\rinmax$, at which the strong-propeller mechanism can work  is related to the mass-flow rate of the disk $\Mdotin$, rotational period $P$, and the dipole moment $\mu$ of the neutron star through  
\be
\Rinmax^{25/8} ~\left|1 -  \Rinmax^{-3/2}\right|
~\simeq ~1.26  ~
\a_{-1}^{2/5} ~M_{1.4}^{-7/6} ~\Mdot_{\m{in,16}}^{-7/20}~ \mu_{30} ~ P^{-13/12} 
\label{1}
\ee  
where $\Rinmax = \rinmax / \rco$,  $M_{1.4}  = (M / 1.4 \Msun)$,  $\Mdot_{\m{in,16}} = \Mdotin / (10^{16} $ \gpers), $\mu_{30} = \mu / (10^{30} $ G cm$^3$), $\a$ is the $\alpha$-parameter of the kinematic viscosity (Shakura \& Sunyaev, 1973), and $\a_{-1} = (\a /0.1)$. 
%%%%%%%%%%%%%%%%%%%%%%%%%%%
\begin{subfigures}
\begin{figure*}
\begin{center}
\vspace{-1.5cm}
\hspace{-1.5cm}
\centering
\includegraphics[width=0.7\textwidth=0.0,angle=-90]{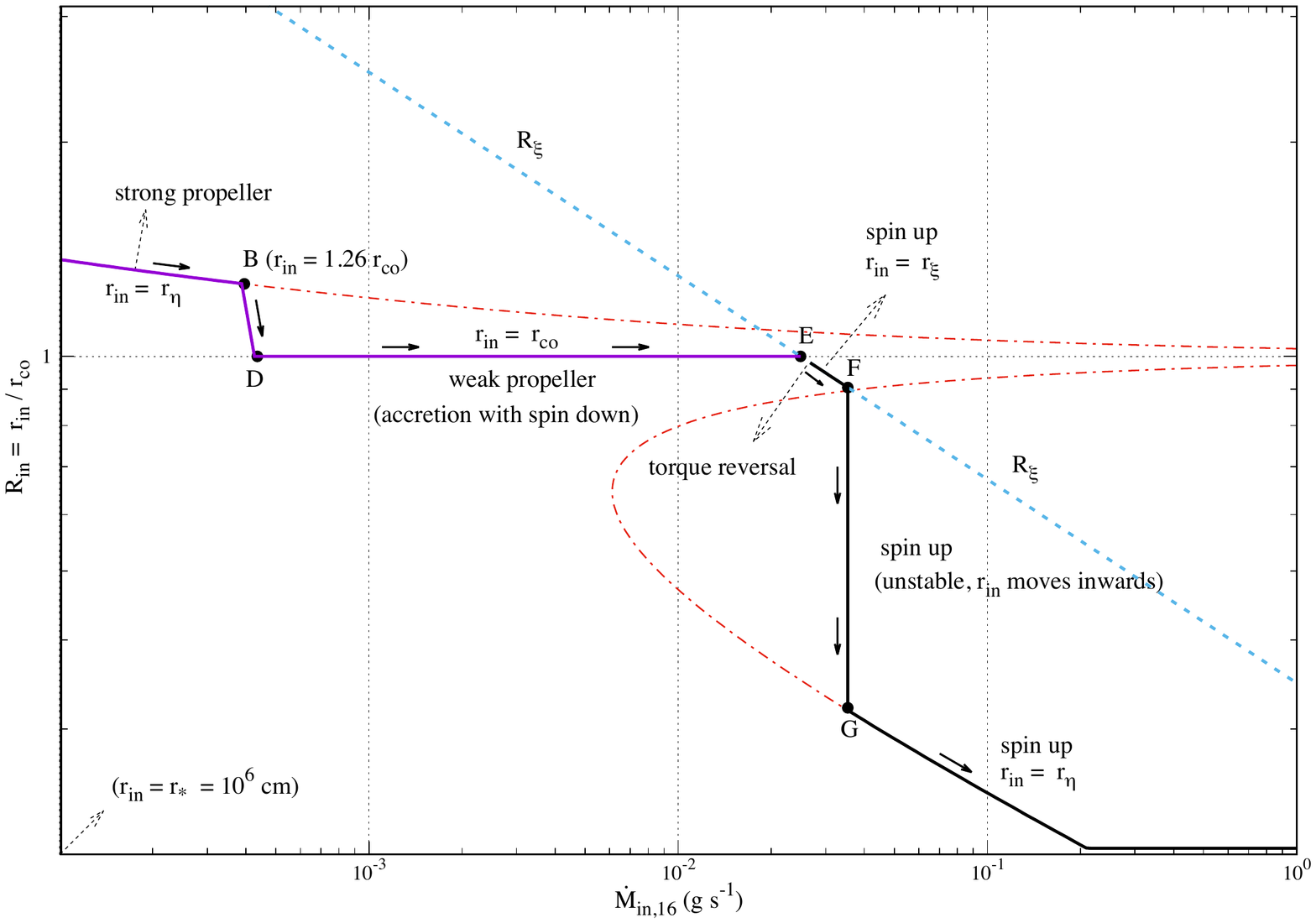}
\vspace{-1.5cm}
\end{center}
\caption{ Variation of $\rin$ with $\Mdotin$ in the strong-propeller phase, the weak-propeller phase, and the spin up phase (solid curve).    Small arrows show the variation of $\rin$ with increasing $\Mdotin$. For this illustrative model, $\Delta r /\rin =0.2$, $\eta= 1.0, \xi = 0.5, B = \mu/ \rstar^3 = 1\times 10^{8}$ G, and $P = 5$ ms (see the text for details).}
\end{figure*}
%%%%%%%%%%%%%%%%%%%%
%%%%%%%%%%%%%%%%%%%%%%%%%%%
\begin{figure*}
\begin{center}
\vspace{-1.5cm}
\hspace{-1.5cm}
\centering
\includegraphics[width=0.7\textwidth=0.0, angle=-90]{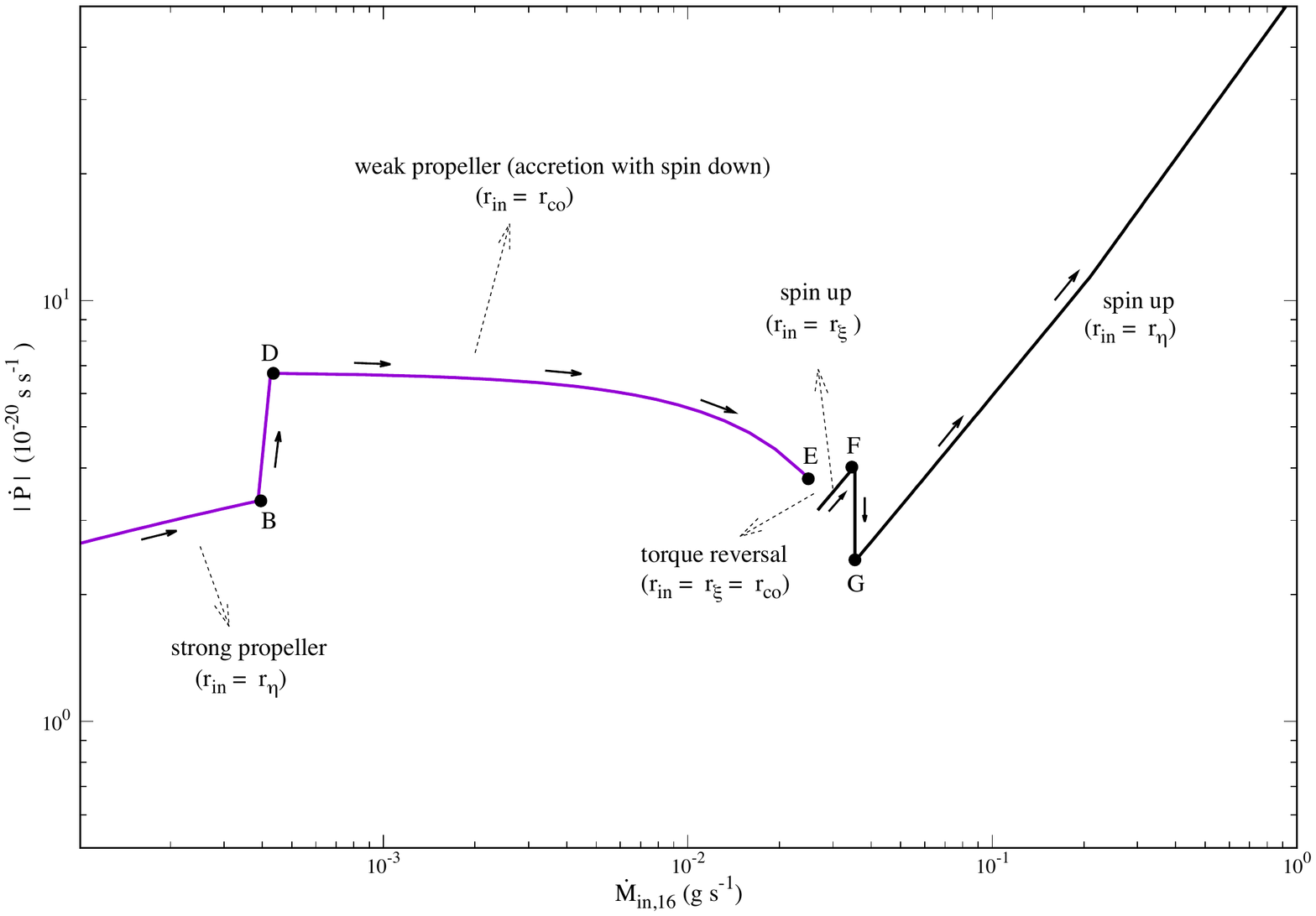}
\vspace{-1.5cm}
\end{center}
\caption{Variation of $\Pdot$ with $\Mdotin$ for the same source given in Fig. 1a. 
The points with letters corresponds to the points in Fig 1a denoted by the same letters.   
Note that the source shows a torque reversal around point E with a small variation in $\Mdotin$. The transition from B to D corresponds to the WP/SP transition with propagation of the inner disk from $r_1$ to $\rco$. For the range of $\Mdotin$ between the points D and E, the system remains in the WP phase with accretion on to the star from $\rin = \rco$ (see the text for details). }  
\end{figure*}
\end{subfigures}
In the SP phase, the inner disk radius is expected to be close to $\rinmax$ because of the sharp radial dependence of magnetic torques.  
We define  the radii $\Reta = \eta \Rinmax $ and $\Rxi = \xi \RA = \xi \rA / \rco$. It can be shown that $\reta$ is always much smaler than $\rA$ in the SP phase (see equation 9 in Ertan 2017).

The solutions for $\Rxi$ (dashed blue curve) and $\Reta$ (dashed dotted red curve)  are given in Fig. (1a) for an illustrative source, a MSP with $P =5$ ms and $\mu = 10^{26}$ G cm$^3$.  As we discussed above, the values of  both $\xi $ and $\eta$ are estimated to be close to unity. For the torque variation of PSR J1023+0038,  a tMSP, Ertan (2018) obtained reasonable results with $\eta \simeq 0.8$.   For all calculations, we set $\xi =0.5$, the value estimated by Ghosh \& Lamb (1979) and $\eta = 1.0$. We will also show the effects of these parameters on our results in Section 3. 
The solid  curve shows the variation of the inner disk radius  $\Rin = \rin / \rco$ (in the direction of the small arrows) with increasing $\Mdotin$. The system is in the SP phase for $\Mdotin$ values up to point B with $\Rin = \Reta$. 
For a given  $\Mdotin$ in this phase, it is seen that $\reta$ is always smaller than $\rxi$. In the conventional models, it is estimated that $\rin = \rxi$,  while  in our model,  $\rin = \reta$ up to the  $\Mdotin$ that gives  $\rin = r_1 = 1.26 \rco$. With increasing $\Mdotin$ beyond this rate,  the system enters the weak-propeller regime  with a transition from B to D (Fig. 1a). The reasons for this distinction between $\rxi $ and $\reta$  are explained in detail by Ertan (2017). The main reason is that all the matter arriving at $\rxi$ cannot be thrown out with speeds greater than $\Uesc$. With the resultant pile-up, the inner disk extends toward inner radii opening up the field lines. Both the field strength and $\tint$ increases with decreasing radius. If the inward motion of the inner disk is stopped at a radius greater than $r_1$, the system settles down to a steady SP regime. This phase could persist  for $\Mdotin $ values up to   $\Mdotin = \Mdotin (\reta = r_1)$. 

\subsection{Weak-propeller (WP) phase} 

With increasing $\Mdotin$ in the strong-propeller phase, if $\reta$ decreases below $r_1$ the inner disk will move gradually inwards until $\rin = \rco$. When $\rin$  is instantaneously between $\rco$  and  $r_1$, the field lines could easily expel matter from the inner boundary because of increasing $\tint$ and  field strength as $\rin$ approaches $\rco$, nevertheless the speed of the outflowing matter cannot exceed $\Uesc$. This leads to a growing pile-up outside $\rin$ which pushes the inner disk inwards until $\rin$ reaches $\rco$ (from B to D in Fig. 1a).  

We estimate that $\rin = \rco$ persists over a large range of  accretion rates (between D and E),  because, in addition to long $\tint$ in the boundary,   viscous stresses are not sufficient for the inner disk to penetrate inside $\rco$ as long as $\rxi$ remains outside $\rco$. The matter couples to the field lines at $\rco$ and flows on to the star along the field lines. In this phase, the conditions inside $\rco$ also force $\rin$  to stay close to $\rco$ (Section 2.3). 
We note that this inner disk behavior is similar to that estimated in the trapped disk 
model (D'Angelo \& Spruit 2012). In our model, the condition for the WP/SP transition and the inner disk radius in the SP phase are well defined, and the physical reason for keeping $\rin = \rco$ in the WP phase is  different from reasons proposed in other models (see also Section 2.3).  

What is the critical accretion rate at which the inner disk can penetrate inside $\rco$? The viscous stresses dominate the magnetic stresses above a critical accretion rate corresponding to  $\rxi = \rco$ (point E in Fig. 1a). 
When the mass inflow rate increases above this level  the inner disk and the boundary region move inwards. During this transition, if the disk-field interaction region remains narrow, as estimated in the simulations (Lovelace et al. 1995), the field lines should decouple from the disk outside $\rco$. This switches off the magnetic spin-down torque produced by the disk-field interaction, leading to the spin-up phase at higher mass inflow rates. The details of the torque calculation are given in Section 2.4. For a range of parameters, our results indicate that the magnetic spin-down torque dominates the spin-up torque produced by accretion on to the star for the entire WP phase. With the disk parameters used in this model, the spin-up is likely to start when the inner disk enters inside $\rco$,  with $\rin = \rxi$  (point E in Fig. 1a).  We will show in Sec. 3 that this result is valid not only for the illustrative  source seen in Fig. 1a, but also for other systems  including strongly magnetized neutron stars with much longer periods, like 4U 1626--67.  

\subsection{Spin-up  (SU) phase}

In the SU phase with $\Rin < 1$, the field lines at the inner boundary should be sufficiently strong to force the matter into co-rotation, like in the SP phase.  In the $\Rin$ calculation for the SU phase, the only difference compared to the SP case is  the sign of $(\Omegastar - \OmegaK)$. This expression appears only in the absolute value in the $\tint$ term (Ertan 2017). This means that equation (\ref{1}) represents  the solutions for the SU phase ($\Rin < 1$) as well. 

For $\Mdotin$ increasing beyond the WP/SU transition rate,  it is seen in Fig. 1a that  the condition $\rin = \rxi$ persist until $\Rxi$ crosses the $\Reta$ curve (between the points E and F). The $\Reta$ solution is double valued for the $\Reta < 1$ region. The positively sloped upper branch appears because of increasing $\tint$ as  $\reta$ approaches $\rco$. This branch of the solution is not stable, but has some important indications: Even for high accretion rates, the inner disk matter can easily be brought into co-rotation at radii close to $\rco$ in the region   between the upper branch of $\Reta$ and $\Rin = 1$  line  (Fig. 1a). In this region, the strength of the field lines are more than sufficient to force the inner disk matter into co-rotation. This also guaranties that  $\rin = \rco$ in the weak-propeller phase when $\rxi > \rco$.  In that case, the high-$\Mdotin$ part of this region is not physically realizable.   For instance, the model source in Fig. 1a can never reach a point  that remains in the high-$\Mdotin$ side of  the E-F line in this region, because $\rin$ cannot be greater than  $\rxi$ where viscous and magnetic stresses are equal.  This is also the reason for the source to follow the E-F line with increasing  $\Mdotin$. 

At  point F, with a slight increase in the accretion rate, the inner disk encounters a range of radii along which the field lines are not able to bring the matter into co-rotation within $\tint$.  This results in an extension of the inner disk inwards, opening the field lines,   down to the radius at which the field is strong enough to sustain the co-rotation.  This stable radius is achieved on the negatively sloped lower branch of the $\reta$ solution at the point corresponding to the same $\Mdotin$ (point G in Fig. 1a). This inner disk radius is stable, and with further increase in $\Mdotin$, the inner disk radius decreases tracking $\reta$.       

The points F and G seen in Fig 1a do not correspond to the same $\Rin$ values for sources with different   $P$ and/or $\mu$.  For some sources,  $\Rxi$ could remain below, and never intersects  the  $\Reta$ curve. In these cases, the inner disk radius remains equal to $\rxi$ for the entire spin-up phase. These possibilities will be discussed with examples in Section 3. 
Due to decreasing $\rin$ with transition from F to G, the torque decreases on the viscous time-scale. We estimate that this sharp torque variation could be observable, if $\Mdotin$ decreases or increases sufficiently slowly through the rate at point F. For some sources,  point G could be found at a radius smaller than the radius of the star, $\rstar$. In this case, when $\Mdotin$ exceeds the rate at point F, the inner disk  extends down to the surface of the star, which is very likely to switch off the pulses produced by the plasma flow channeled  by the field lines on to the poles. A model source to illustrate this situation is discussed in Section 3.    
 
\subsection{Torques}

The total torque acting on the star could be written as the sum of spin-up and the spin-down torques

\be
\Gamma = ~\Mdotstar~ (G M \rin)^{1/2} 
- ~\frac {\mu^2}{\rin^3} \left(\frac{\Delta r}{\rin}\right)  + \Gamma_{\m{dip}} 
\label{4}
\ee 
 The first term is the spin-up torque associated with accretion on to the star from the inner disk. The second term is the magnetic torque arising from the disk-field interaction inside the boundary with radial width $\Delta r$.  We take $\Delta r/r = 0.2$ in all calculations. Since in all cases we consider here, this magnetic torque dominates the dipole radiation torque,  $\Gamma_{\m{dip}}$, we have ignored   $\Gamma_{\m{dip}}$ in our calculations.

In the strong-propeller phase, there is no accretion on to the star, that is, the first term is zero, $\rin = \reta = \eta \rinmax$. 
In the weak-propeller phase,  both the accretion torque and the magnetic torque act on the star  with $\rin = \rco$. In the SU phase, only the accretion torque is active since the inner disk boundary interacting with the field lines is expected to lie inside the co-rotation radius.
In this phase, we take $\rin = \rxi$ until $\rxi$ crosses $\reta$ (between the points E and F), and $\rin = \reta$ for the accretion rates greater than the rate at point G  (see Fig. 1b).

\begin{subfigures}
\begin{figure}
%\begin{center}
\vspace{-1.5cm}
\hspace{-2.2 cm}
%\centering
\includegraphics[trim=10cm 0 0 -1cm, width=8.0cm,angle=-90]{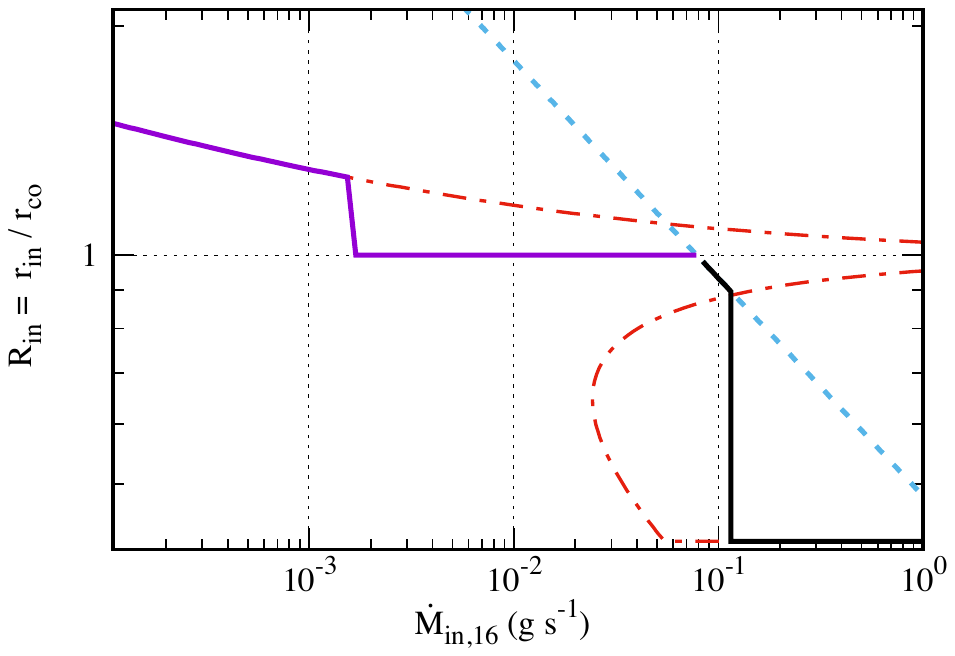}
\hspace{-2.2 cm}
\vspace{-1.8 cm}
%\end{center}
\caption{The same as Fig.1a, but with $B = 5\times 10^7$ G, $P$ = 1.69 ms (properties of PSR J1023+0038).  }
%\label{Fig 3a} 
\end{figure}

\begin{figure}
%\begin{center}
\vspace{-1.7cm}
\hspace{-2.2 cm}
%\centering
%\includegraphics[width=0.7\textwidth=0.0,angle=0]{fig1.pdf}
%\includegraphics[width=0.8\textwidth=0.0,angle=-90]{mdot2.pdf}
\includegraphics[trim=10cm 0 0 -1cm, width=8.0cm,angle=-90]{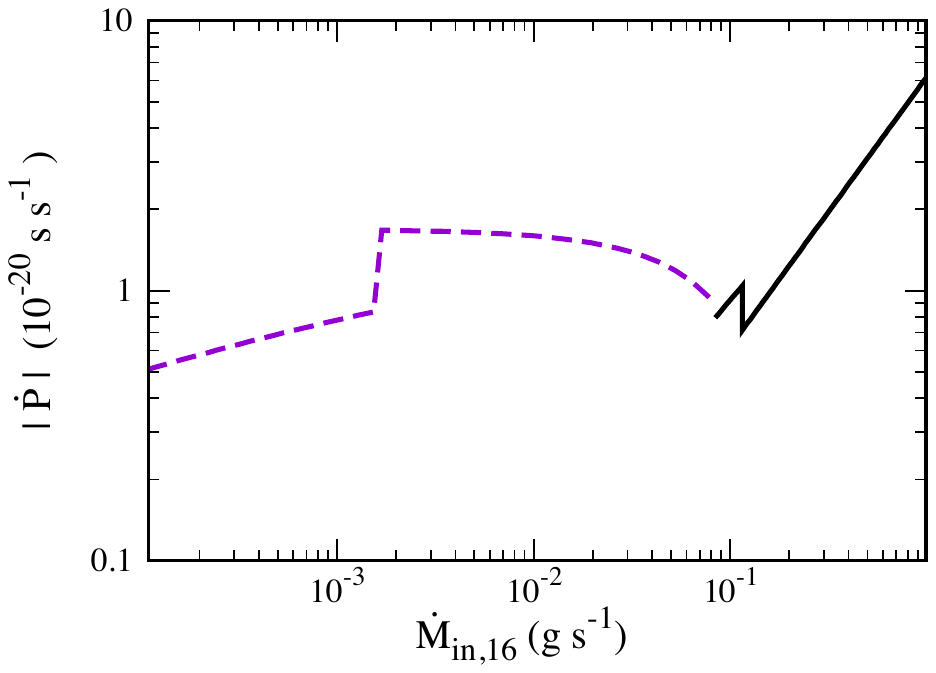}
\hspace{-2.2 cm}
\vspace{-1.8cm}
%\end{center}
\caption{$|\Pdot|$ curve for the same source given in Fig 2a.}
%\label{fig 3b} 
\end{figure}
\end{subfigures}

We obtain  the $\Pdot$ curve given in Fig. 1b  for the same source with the $\Mdotin$ curve given in Fig 1a. The points corresponding to the same $\Mdotin$ values in Figs. 1a and 1b are denoted with the same letters.  The sharp increase in $\Pdot$ from B to D is due to the inward motion of the inner disk from $r_1 = 1.26 \rco$ to $\rco$ with a slight increase in $\Mdotin$ when $\rin = \reta = r_1$  (Section 2.2).   With this transition from the SP phase to the WP phase accretion on to the star is switched on.    

The system is in the WP regime between points D and E. In this phase, $\rin = \rco$ as explained in Sec. 2.2. The field interacts with the inner disk in the narrow boundary just outside $\rco$, which exerts a spin-down torque  on the star. The mass-flow from $\rin = \rco$ along the field lines on to the poles causes a spin-up torque. For the parameters given in Fig. 1a, the spin-down torque dominates the spin-up torque for the entire WP phase. As seen in Fig. 1b, the magnitude of the spin-down torque is decreasing with increasing $\Mdotin$ toward the point E. This is because the magnitude of the accretion torque becomes comparable to, while still less than, the magnetic spin-down torque as $\rxi$ approaches $\rco$.  In Sec. 3, we also discuss how the features of this transition depend on the chosen parameters $\Delta r /\rin, \eta$ and $\xi$ , as well as $\mu$ and $P$. 
 
The inner disk penetrates inside $\rco$ when $\rxi$  becomes smaller than $\rco$,  taking the star from the WP to the SU phase. 
For given $\mu$ and $P$, the critical $\Mdotin$ corresponding to this transition can be estimated by equating $\rxi$ to $\rco$. 
The sharp decrease in the magnitude of the torque from  F to G corresponds to the propagation of $\rin$ from the unstable point F to the stable point G on the lower branch of the $\reta $ curve (see Figs. 1a and 1b). With further increase in $\Mdotin$,  the inner disk radius $\rin = \reta$ decreases while  the magnitude of the spin-up torque increases as seen in Fig. 1b.

\begin{subfigures}
\begin{figure}
%\begin{center}
\vspace{-1.5cm}
\hspace{-2.2 cm}
%\centering
%\includegraphics[width=0.7\textwidth=0.0,angle=0]{fig1.pdf}
%\includegraphics[width=0.8\textwidth=0.0,angle=-90]{mdot2.pdf}
\includegraphics[trim=10cm 0 0 -1cm, width=8.0cm,angle=-90]{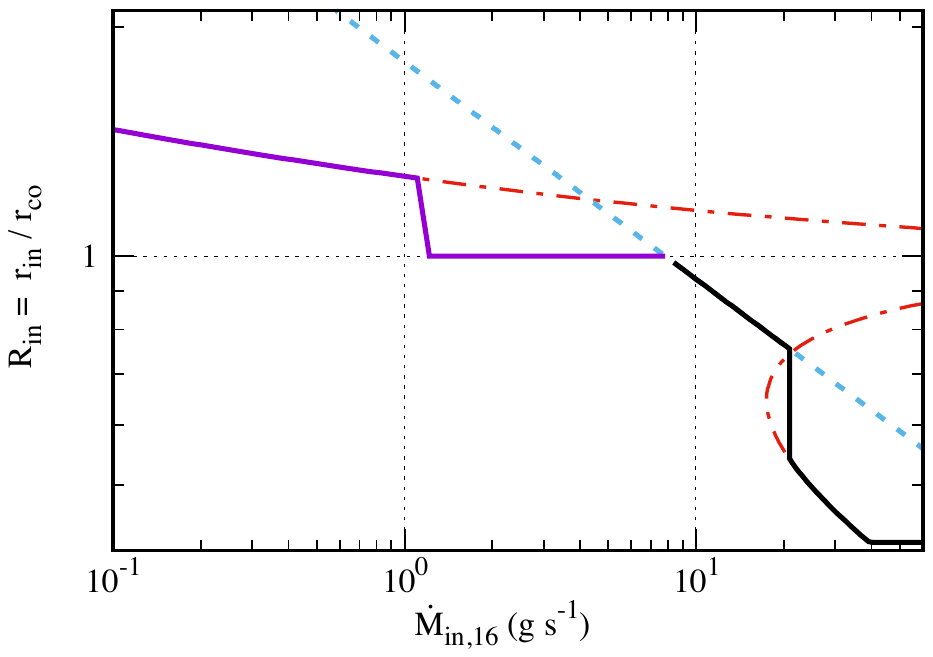}
\hspace{-2.2 cm}
\vspace{-1.8cm}
%\end{center}
\caption{The same as Fig.2a, but with$B = 5 \times 10^8$ G}
%\label{Fig 3a} 
\end{figure}

\begin{figure}
%\begin{center}
\vspace{-1.7cm}
\hspace{-2.2 cm}
%\centering
\includegraphics[trim=10cm 0 0 -1cm, width=8.0cm,angle=-90]{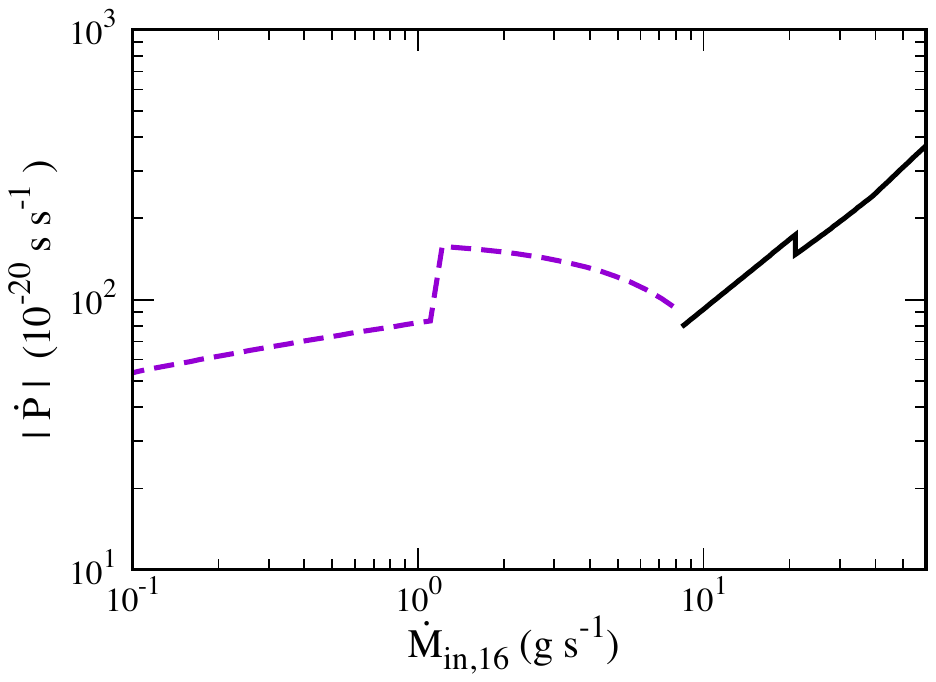}
\hspace{-2.2 cm}
\vspace{-1.8cm}
%\end{center}
\caption{$|\Pdot|$ curve for the same source given in Fig 3a.}
%\label{fig 3b} 
\end{figure}
\end{subfigures}

\begin{subfigures}
\begin{figure}
%\begin{center}
\vspace{-1.5cm}
\hspace{-2.2 cm}
%\centering
%\includegraphics[width=0.7\textwidth=0.0,angle=0]{fig1.pdf}
%\includegraphics[width=0.8\textwidth=0.0,angle=-90]{mdot2.pdf}
\includegraphics[trim=10cm 0 0 -1cm, width=8.0cm,angle=-90]{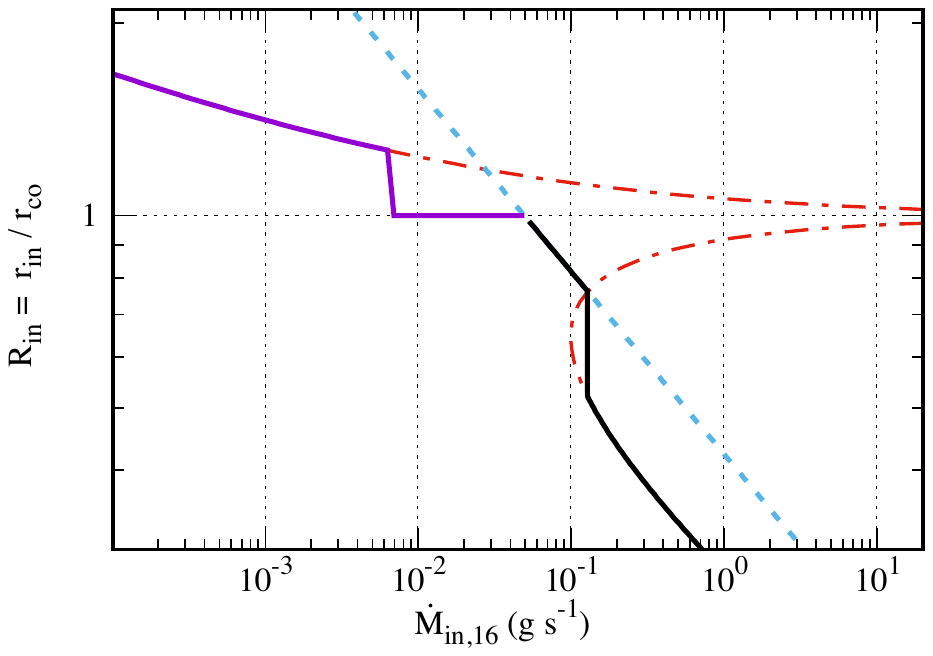}
\hspace{-2.2 cm}
\vspace{-1.8cm}
%\end{center}
\caption{The same as Fig.1a, but with $B = 1 \times 10^{12}$ G, $P$ = 10 s}
%\label{Fig 3a} 
\end{figure}

\begin{figure}
%\begin{center}
\vspace{-1.7cm}
\hspace{-2.2 cm}
%\centering
%\includegraphics[width=0.7\textwidth=0.0,angle=0]{fig1.pdf}
%\includegraphics[width=0.8\textwidth=0.0,angle=-90]{mdot2.pdf}
\includegraphics[trim=10cm 0 0 -1cm, width=8.0cm,angle=-90]{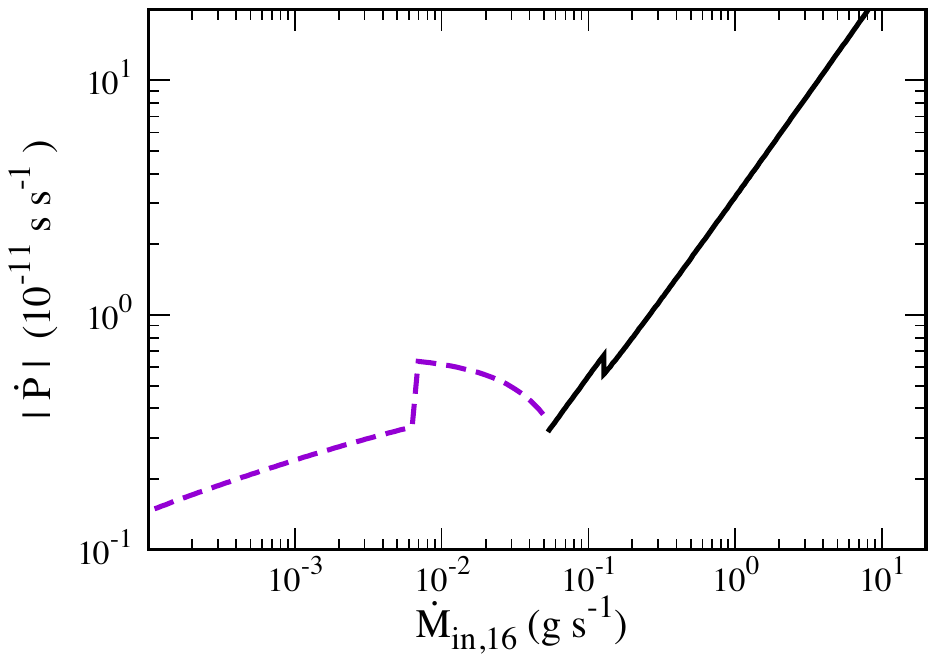}
\hspace{-1.7 cm}
\vspace{-1.8cm}
%\end{center}
\caption{$|\Pdot|$ curve for the same source given in Fig 4a.}
%\label{fig 3b} 
\end{figure}
\end{subfigures}

\begin{subfigures}
\begin{figure}
%\begin{center}
\vspace{-1.2cm}
\hspace{-2.2 cm}
%\centering
%\includegraphics[width=0.7\textwidth=0.0,angle=0]{fig1.pdf}
%\includegraphics[width=0.8\textwidth=0.0,angle=-90]{mdot2.pdf}
\includegraphics[trim=10cm 0 0 -1cm, width=8.0cm,angle=-90]{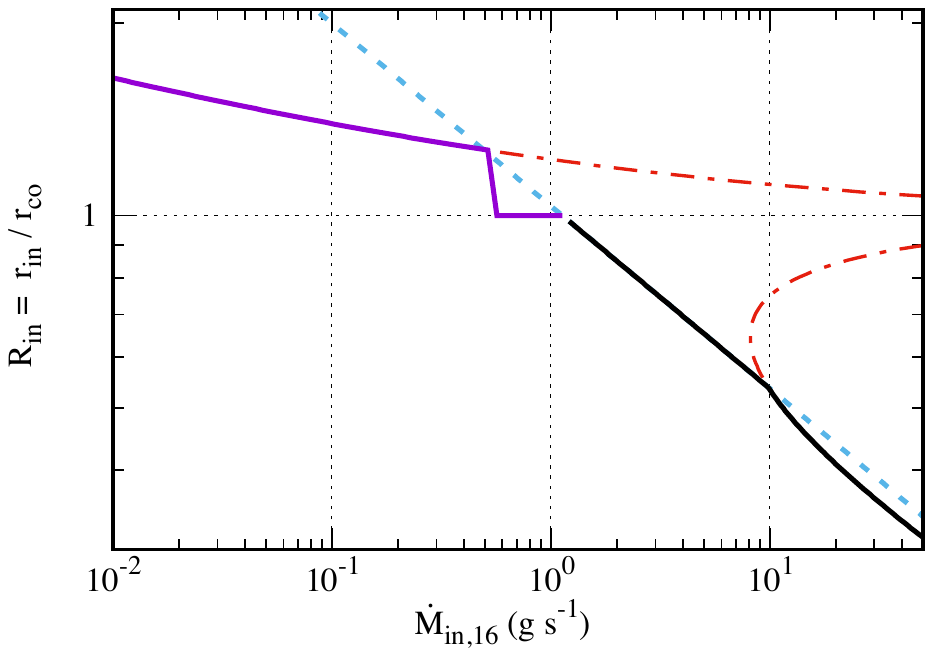}
\hspace{-2.2 cm}
\vspace{-1.8cm}
%\end{center}
\caption{The same as Fig.1a, but with $B = 3.5 \times 10^{12}$ G, $P$ = 7.66 s, the properties of 4U 1626--67}
%\label{Fig 3a} 
\end{figure}

\begin{figure}
%\begin{center}
\vspace{-1.7cm}
\hspace{-2.2 cm}
%\centering
%\includegraphics[width=0.7\textwidth=0.0,angle=0]{fig1.pdf}
%\includegraphics[width=0.8\textwidth=0.0,angle=-90]{mdot2.pdf}
\includegraphics[trim=10cm 0 0 -1cm, width=8.0cm,angle=-90]{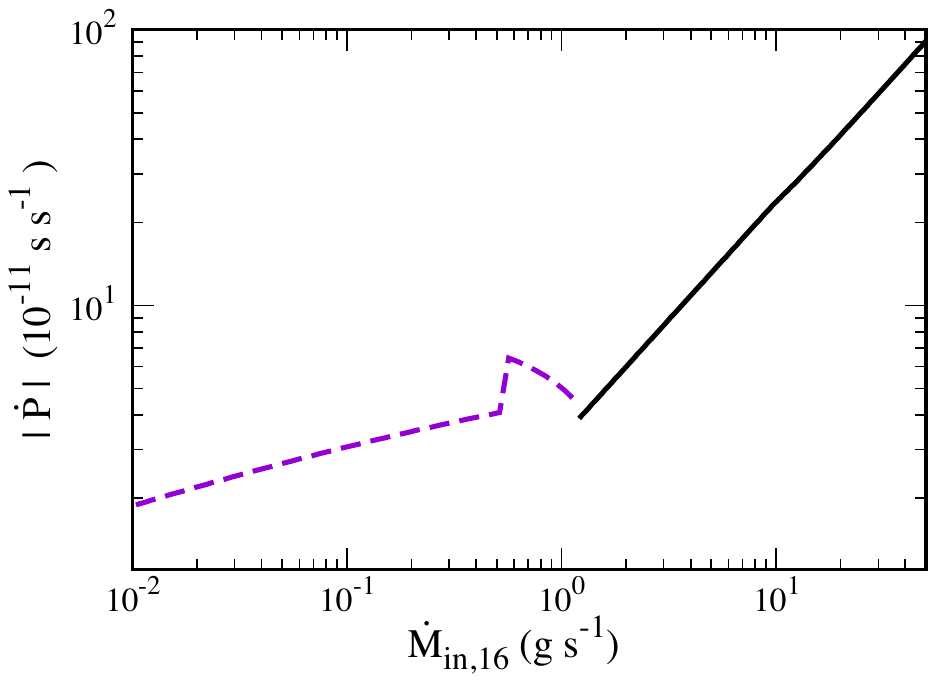}
\hspace{-2.2 cm}
\vspace{-1.8cm}
%\end{center}
\caption{$|\Pdot|$ curve for the same source given in Fig 5a.}
%\label{fig 3b} 
\end{figure}
\end{subfigures}

The model curves given in Figs (1-5) are obtained with the the same disk parameters ($\xi, \eta, \Delta r / \rin$). The critical  $\Mdotin$ and $\rin$ values corresponding to the SP/WP  and torque-reversal transitions change depending on $\mu$ and $P$ of the sources. 
The $\mu$ and $B$ values producing the model curves in Figs. (1-3)  and Figs. (4-5) are typical values for the millisecond pulsars and LMXBs (or HMXBs) like 4U 1626--67 respectively. Tracing orders of magnitude ranges for $\mu$ and/or $P$ values, we always find the same torque-reversal behavior: an abrupt change in the sign of the torque without a significant change in its magnitude occurs with a small change in the mass-flow rate of the disk.

\section{Discussion}

The model curves seen in Figs. (2 - 5) are obtained with the same set of parameters used for the model given in Fig. 1, except for the $B = \mu / \rstar^3$ and $P$ values. For the source seen in Fig. 2, $P=1.69$ ms and $B = 5 \times 10^{7}$ G,  the properties of   tMSP PSR J1023+0038 are employed. In Fig. 2a, it is seen that systems similar to this source are not expected to show pulsations at accretion rates greater than about $10^{15}$  \gpers.  Comparing with Fig. 1a, it is seen  in Fig. 2a that the radius at point G  is smaller than $\rstar$, indicating that the inner disk extends to the surface of the star. For this source, $\rstar = 1 \times 10^6$ cm corresponds to $\Rin = 0.42$.    This model source shows the WP/SU transition  at  slightly lower $\Pdot$ and $\Mdotin$ values in comparison with the source given in Fig. 1. 

For the model curve given in Fig. 3 all the parameters are the same as those given in Fig. 2, except the field strength is increased to $B= 5 \times 10^{8}$ G, which shifts the critical $\Mdotin$ values upwards for both the SP/WP transition and the torque reversal. The critical torque reversal rate increases because $\rxi = \rco$ requires a greater $\Mdotxi$ for a stronger field ($\rxi \p B^{4/7} \Mdotin^{-2/7}$).   The magnitudes of the spin-up and spin-down torques close to the torque reversal also increase with increasing $B$, while their ratio  remains similar. For this case, the crossing point F is at a slightly smaller $\Rin$,  while the points F and G are closer to the minimum of  $\reta$ curve. 

The source given in Fig. 4, illustrates a strongly magnetized neutron star with $B= 1 \times 10^{12}$ G  and $P = 10$ s, parameters typical of HMXBs or LMXBs like 4U 1626--67, and very different from those of the millisecond pulsars discussed above. It is seen, by comparing Figs. 2 and 4, that the torque reversal for these very different systems takes place at similar  $\Mdotin$, but with orders of magnitude different $\Pdot$ values. It is very remarkable that the ratios of the of the spin-up and spin-down torque magnitudes across the torque reversals are similar for these rather different model sources.

\begin{subfigures}
\begin{figure}
%\begin{center}
\vspace{-1.5cm}
\hspace{-2.2 cm}
%\centering
%\includegraphics[width=0.7\textwidth=0.0,angle=0]{fig1.pdf}
%\includegraphics[width=0.8\textwidth=0.0,angle=-90]{mdot2.pdf}
\includegraphics[trim=10cm 0 0 -1cm, width=8.0cm,angle=-90]{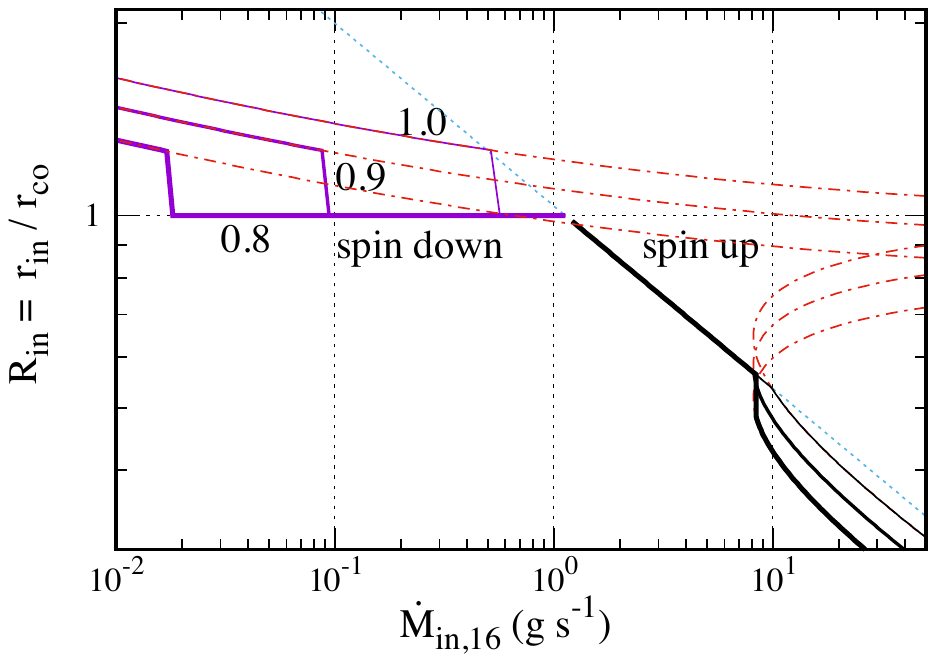}
\hspace{-2.2 cm}
\vspace{-1.8cm}
%\end{center}
\caption{The same as Fig.5a, but with different $\eta$ values seen on the model curves }
%\label{Fig 3a} 
\end{figure}

\begin{figure}
%\begin{center}
\vspace{-1.7cm}
\hspace{-2.2 cm}
%\centering
%\includegraphics[width=0.7\textwidth=0.0,angle=0]{fig1.pdf}
%\includegraphics[width=0.8\textwidth=0.0,angle=-90]{mdot2.pdf}
\includegraphics[trim=10cm 0 0 -1cm, width=8.0cm,angle=-90]{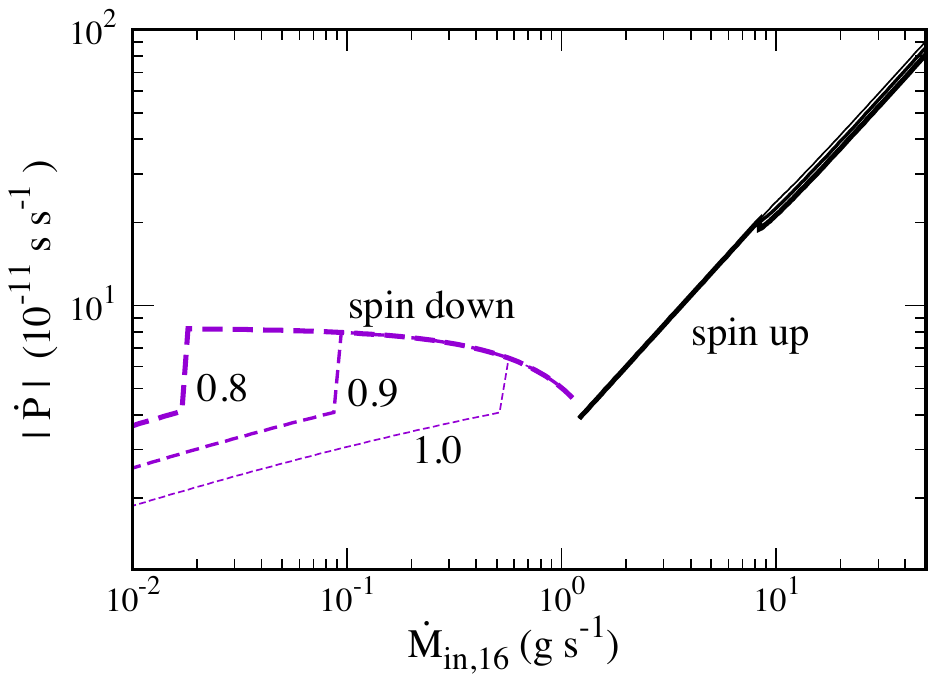}
\hspace{-2.2 cm}
\vspace{-1.8cm}
%\end{center}
\caption{$|\Pdot|$ curves for the sources given in Fig 6a.}
%\label{fig 3b} 
\end{figure}
\end{subfigures}

The model curves in Fig. 5 are produced with the estimated properties of 4U 1626--67, $B = 3.5\times 10^{12}$ G and $P = 7.66$ s  (Giacconi at al. 1972, Orlandini et al. 1998). For this source, the  $\Mdotin$ corresponding to the torque reversal is about $ 10^{16}$ \gpers.  As seen in Fig. 5b the magnitudes of the spin-up torque and and the spin-down torque are similar around the torque reversal with $|\Pdot| \simeq 3 \times 10^{-11}$ \spers  ~($|\dot{\nu}| \simeq 5 \times 10^{-13}$ Hertz s$^{-1}$). This is consistent with the observed rotational rates of the source during the torque reversals  (Camero-Arranz et al. 2010, Chakrabarty  et al. 1997a). The torque reversal could occur with a small change in $\Mdotin$. 

The model parameter $\eta$ does not affect the  WP/SU transition, while it changes the critical $\Mdotin$ for the SP/WP transition. The model curves seen in Fig. 6 illustrate the effect of $\eta$ on these transitions. 
The effects of the parameters $\xi$ and $\Delta r / \rin$ on the ratio of the torque magnitudes during the torque reversals are degenerate.  For a  smaller $\xi = \rxi / \rA$, the transition  takes place at a lower  $\Mdotin$ with  a smaller accretion torque, while a smaller $\Delta r / \rin$  gives a weaker spin-down torque during the torque reversal. To illustrate this effect, we plot the model curve in Fig. 7, using the same parameters of the model curve with $\eta = 0.9$ in Fig.6, except with a greater $\xi$ parameter ($\xi = 0.7$). As seen in Fig. 7, the torque reversal with incresing $\Mdotin$ could take place before the inner disk penetrates into $\rco$. Like in the previous examples, torque reversal occurs with a small change in  $\Mdotin$, and with similar torque magnitudes before and after the torque reversal. In Fig. 8, we plot the torque variations for three different $\Delta r / \rin$ values.  It is seen that the spin-down torque is increasing with increasing  $\Delta r / \rin$, while the spin-up torque remains constant, since this parameter does not affect the accretion torque.  

For the illustrative sources, with different $P$ and $\mu$ values,  Figs. 1-5 show that transitions occur without significant changes  in $\Mdotin$ and $|\Pdot|$, which are in agreement with observed torque reversal properties (Section 1).  
Considering that the disk parameters $\eta, \xi$ and  $\Delta r / \rin$ are roughly similar for different systems, the ratio of the spin-down and spin-up torques across the WP - SU torque reversal transition is expected to be similar for different sources, and for a reasonable set of disk parameters our results indicate that this ratio is likely to be close to unity. The timescale for the torque reversal depends on the rate of change of $\Mdotin$ during the transition.  

\begin{subfigures}
\begin{figure}
%\begin{center}
\vspace{-1.5cm}
\hspace{-2.2 cm}
%\centering
%\includegraphics[width=0.7\textwidth=0.0,angle=0]{fig1.pdf}
%\includegraphics[width=0.8\textwidth=0.0,angle=-90]{mdot2.pdf}
\includegraphics[trim=10cm 0 0 -1cm, width=8.0cm,angle=-90]{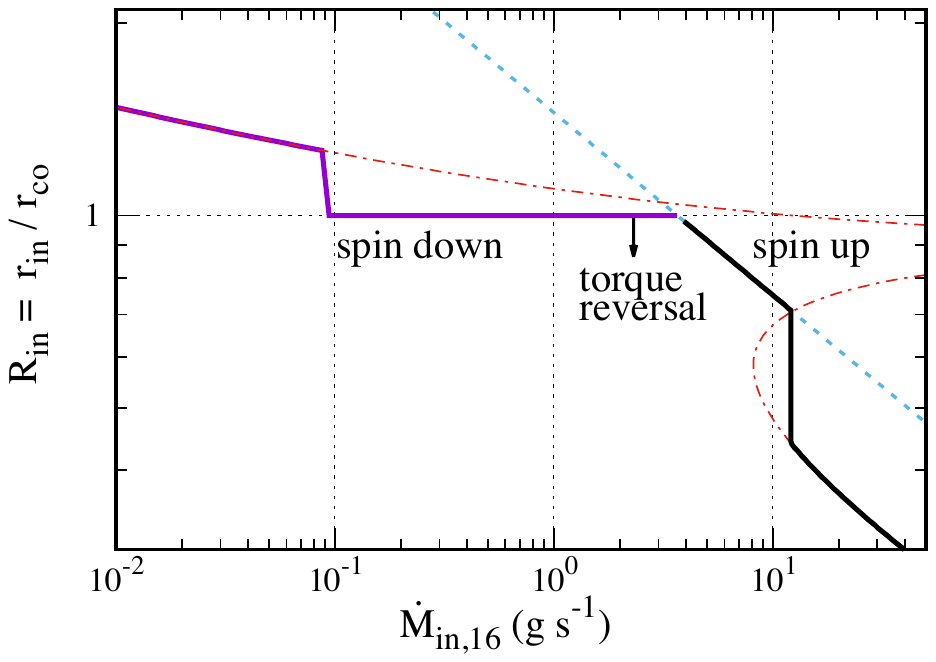}
\hspace{-2.2 cm}
\vspace{-1.8cm}
%\end{center}
\caption{The same as the curve in Fig.6 with $\eta = 0.9$, but with $\xi = 0.7$.  }
%\label{Fig 3a} 
\end{figure}
\begin{figure}
%\begin{center}
\vspace{-1.7cm}
\hspace{-2.2 cm}
%\centering
%\includegraphics[width=0.7\textwidth=0.0,angle=0]{fig1.pdf}
%\includegraphics[width=0.8\textwidth=0.0,angle=-90]{mdot2.pdf}
\includegraphics[trim=10cm 0 0 -1cm, width=8.0cm,angle=-90]{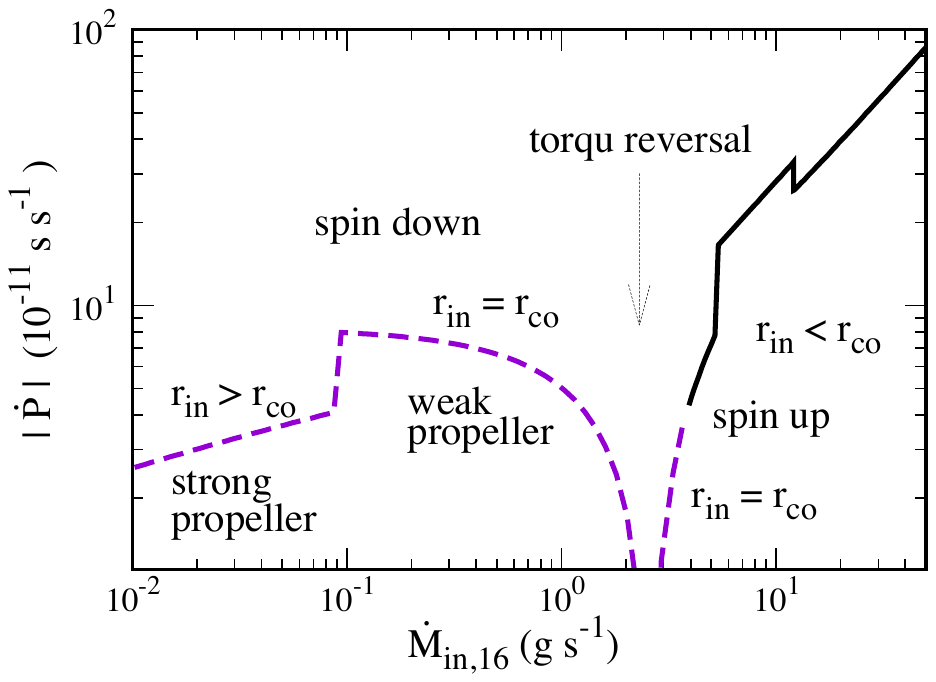}
\hspace{-2.2 cm}
\vspace{-1.8cm}
%\end{center}
\caption{$|\Pdot|$ curve for the source given in Fig 7a.}
%\label{fig 3b} 
\end{figure}
\end{subfigures}

When the system is in the weak-propeller phase, small variations in  $\Mdotin$ could lead to occasional transitions between the WP and SP phases with durations in both states much longer than the viscous time-scale of the inner disk. 
In our model, the $\Mdotin$ dependence of $\reta$ in the strong-propeller phase is much weaker than that of commonly assumed inner disk radius $\rxi$ (see Fig. 1a). This means that once $\rco < \rin < r_1$, it is not easy to refill the inner disk due to this behavior of $\reta$. 
Both $B$ and $\tint$ increases as $\rin$ comes close to $\rco$ in the strong-propeller phase. For a steady mass-flow from the outer disk, 
if $\rco < \rin < r_1$ instantaneously, it is inevitable that $\rin$ will decrease, and eventually become equal to $\rco$. This requires a significant growth of the inner pile-up while the matter is being expelled from the inner boundary to larger radii continuously,  which could take a time much longer than the viscous time-scale.
Note that the weak $\Mdotin$ dependence of $\reta$ also guaranties that $\rin$ remains close to $\rco$ in this transient strong-propeller phase. 
The details of these events including recurrence times should be studied through numerical analysis. We note here that these events could lead to transitions between the radio pulsar and the LMXB states, and between the high X-ray modes (X-ray pulsar) and the low X-ray modes (no X-ray pulses)  of tMSPs (Ertan 2018).

\begin{figure}
%\begin{center}
\vspace{-1.7cm}
\hspace{-2.2 cm}
%\centering
%\includegraphics[width=0.7\textwidth=0.0,angle=0]{fig1.pdf}
%\includegraphics[width=0.8\textwidth=0.0,angle=-90]{mdot2.pdf}
\includegraphics[trim=10cm 0 0 -1cm, width=8.0cm,angle=-90]{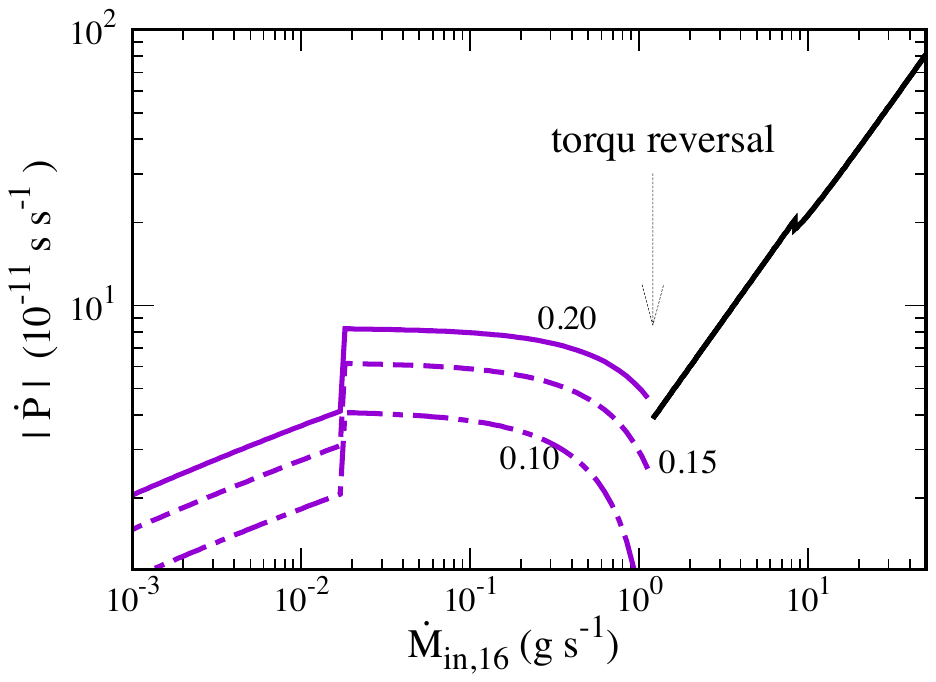}
\hspace{-2.2 cm}
\vspace{-1.8cm}
%\end{center}
\caption{$|\Pdot|$ curves for different $\Delta \rin / r$ values seen on the model curves for $B = 3.5 \times 10^{12}$ G, $P$ = 7.66 s., $\eta =0.8$}
%\label{fig 3b} 
\end{figure}

In the weak-propeller phase, the inner disk is not likely to significantly penetrate inside $\rco$ when $\Rxi > 1$. Because, in addition to long $\tint$ in the boundary region, the $\reta$ solution for the spin-up phase implies that the field lines inside and close to $\rco$ can easily bring the matter into co-rotation forcing  accretion on to the star from $\rco$ for the accretion rates lower than  $\Mdotin (\Rxi = 1)$  (see Section 2).  To sum up, in this phase, due to the conditions around $\rco$, the inner disk radius always tends to be very close to $\rco$ between the critical accretion rates defined in Section 2. The field lines interacting with the inner disk could produce a continuous backflow of matter from the inner boundary to larger radii, and also injects  angular momentum to the pile-up which is the dominant mechanism generating the spin-down torque. 

Finally, before and after the transitions from the SU to the WP phase with some rapid changes in the accretion rate, the inner disk may not immediately establish the steady-state conditions. Possibly, there could also be small variations in the geometry of the inner boundary with changing $\Mdotin$ as well. These effects, which are not addressed in our model, could lead to torque and X-ray luminosity variations in ways different from expected in our  model for a particular accretion regime. 
For instance, the gradual decrease in the spin-down rate of 4U 1626--67 following its transition to the spin-down phase (Chakrabarty et al. 1997a) by $\sim 30 \%$  accompanying the decrease in the X-ray flux by a factor of $\sim 2$  (Camero-Arranz et al. 2010)  could be due to these $\Mdotin$ dependent small variations in the inner boundary. As pointed out by Bildsten et al. (1997), the change in the observed X-ray flux in a particular energy band should also be taken with some caution, considering that it may not reflect the variation in the bolometric luminosity.

Our results in this work are obtained for the neutron stars with geometrically thin, optically thick accretion disks. Wind accretion could significantly change the torques and the critical condition for the torque reversals. In particular, observations indicate a persistent accretion disk in LMXB GX 1+4 (Chakrabarty \& Roche 1997). In this wide binary system (Hinkle et al. 2001), Roche-lobe overflow is unlikely. The source of mass-flow on to the neutron star is estimated to be slow stellar wind from the giant companion forming  a quasi-spherical  accretion geometry (Gonz$\acute{\mathrm{a}}$lez-Gal$\acute{\mathrm{a}}$n et al. 2012) for which  the torque-reversal conditions cannot be estimated in our model. Another disk-fed system Cen X-3 (Bildsten et al. 1997) is an HMXB. Its disk is likely to form through Roche-lobe overflow, while the X-rays are estimated to be obscured occasionally by the wind of the companion (Camero-Arranz et al.  2010). Superimposed on a long-term secular spin-up with $\dot{\nu} \sim 8 \times 10^{-13}$ Hz s$^{-1}$,  the source shows rapid transitions between the steady spin-up and spin-down phases lasting 10 - 100 days with average  $\dot{\nu}$ magnitudes of   $\sim 7 \times 10^{-12}$ Hz s$^{-1}$ and $\sim 3 \times 10^{-12}$ Hz s$^{-1}$ respectively (Bildsten et al. 1997). If the disk-field interaction is the dominant torque mechanism acting on this source,  a dipole moment $\mu_{30} \sim 5 $ with $\Mdotin \simeq 1\times10^{17}$  \gpers produces a torque reversal in our model consistently with the observed torques. Considering that the wind effect may not be negligible, this result should be taken with some caution.

Observations of  4U 1626--67  clearly showed that the source has a geometrically thin (optically thick ) disk  before and after the torque reversals (Camero-Arranz et al. 2010). The emission line complex around 1 keV and iron K-fluorescence line at 6.4 keV observed in both the spin-up and the spin down phases show that the temperature of the inner disk environment remained below a few keV, which implies that the optical structure of the disk did not change during the torque reversal (Camero-Arranz et al. 2012). Furthermore, the accretion rates estimated to be greater than about $10^{15}$ \gpers from the X-ray spectra for the both torque reversals are also in agreement with the presence of a geometrically thin disk around the source in both spin-up and spin-down. During the torque reversals, there are also systematic changes in the spectra (Camero-Arranz et al. 2012) and pulse profiles that remain stable in the long-lasting spin-up and spin-down phases (Beri et al. 2014). In our model,  the inner disk penetrates into  $\rco$ and propagates inwards after the transition to the spin-up phase. This could result in significant changes in the geometry and optical properties of the accretion column of the star, and thereby, also change the pulse profile, emission area and temperature, while the emission line properties are modified by the newly established conditions at the inner disk. Nevertheless,  the details of these variations in the spectra, pulse shapes and the line properties are not addressed in our model. 

When this work was in preparation, we noticed the detailed model fits to the torque reversal data of 4U 1626--67 by Benli (2020)  assuming  $\rin = \rco$ in both spin-up and spin-down phases.  The results with this assumption correspond to the particular case in this work given in Fig. 7 for which the inner disk cannot enter inside $\rco$ even in the spin-up phase below a critical accretion rate. In our model, with increasing mass-flow rate above this critical level, $\rin$ moves inwards as described in Section 2.  The results obtained by Benli (2020) seem to favor this torque-reversal behavior. We will perform detailed model fits to the torque reversal data of the sources in an independent work.

\section{ SUMMARY and CONCLUSIONS}

Extending our earlier work  (Ertan 2017, 2018), we have presented a complete picture  that could explain the basic properties of neutron stars accreting from geometrically thin accretion disks in the three main states, namely the strong-propeller (SP) phase, the weak-propeller (WP)  phase, and the spin-up (SU) phase. In the SP phase, the inner disk radius, $\rin$, is much smaller than the conventional \Alfven radius, $\rA$, and greater than $r_1 = 1.26 \rco$. In this phase, all the inflowing disk matter is expelled from the system. The star spins-down by the magnetic torques arising from the disk-field interaction.  With increasing $\Mdotin$, the system makes a transition from the SP to the WP phase when $\rin$ decrease below $r_1$, and eventually reaches $\rco$. During this transition, the magnitude of the net spin-down torque decreases due to the contribution of the spin-up torque switched on with the onset of accretion when  $\rin = \rco$.  In the WP phase, we estimate that most of the inflowing mass is accreted on to the neutron star and the star spins down, since the magnetic torque dominates the spin-up torque for a large range of accretion rates. The inner disk penetrates into $\rco$ only when viscous stresses dominate the magnetic stresses at  $\rco$, which requires  $\rxi = \xi \rA < \rco$.  This switches off the magnetic spin-down torque and starts the SU phase. The inner disk with $\rin = \rxi$ during the WP/SU transition, propagates inwards, with increasing accretion rate, to radii smaller than $\rxi$.        
Depending on the actual values of inner disk parameters, the torque reversal could take place before the inner disk penetrate into $\rco$ (Fig. 7).     

The model can account for: (1)  accretion on to the neutron star at low X-ray luminosities, (2) transition from the SP to the WP phase at accretion rates much lower than the rates corresponding to the spin-up/spin down transition, (3) ongoing accretion with spin-down over a large range of X-ray luminosity, 
and (4) torque reversals with comparable torque magnitudes without a significant change in the accretion rate.   
We have elaborated on this behavior in the context of the tMSPs and extended them to the source 4U 1626-67  which has a much stronger dipole field  and a persistent accretion disk, and show torque reversals not affected by the wind of the companion. There are some other disk-fed LMXB and HMXB systems  that show torque reversals with similar properties (Bildsten et al. 1997). Since their torque-luminosity relations are not clear due to either the effect of the wind from the companion or the inconvenient viewing  geometry (see Section 1), one can not study their torque-reversal properties. We have presented illustrative model curves for different source properties, which can be tested through future observations of both SP/WP transitions and the torque reversals.    

Results of our model calculations indicate that the critical accretion rate corresponding to the torque reversal depends on $P$ and $\mu$ of the source. We find that the magnitudes of torques before and after the torque reversal are similar with a ratio that remains almost the same for systems with  different $P$ and $\mu$ values.  Our results also show that the torque reversals could occur without a significant change in the accretion rate independently of  the torque-reversal luminosity, period and dipole moment of the sources. In our model, the main reason for the abrupt torque reversal with similar torque ratios  is that the inner disk radius is cut at $\rco$ in the WP phase, and for the torque model we employ, magnitudes of the net torques in either side of the torque reversal are always found to be similar independent of  $\mu$ and $P$  (Figs. 1-5). Exact values of the critical $\Mdotin$  for the torque reversals depend on the actual values of the disk parameters ($\xi, \eta, \Delta \rin / r$). Since these parameters are likely to be similar for different sources, our results imply that the ratios of the spin-up and spin-down torques around the torque reversals are likely to be similar for different neutron stars accreting from geometrically thin accretion disks.


\section*{Acknowledgements}

We acknowledge research support from
T\"{U}B{\.I}TAK (The Scientific and Technological Research Council of
Turkey) through grant 117F144 and from Sabanc\i\ University. We thank Ali Alpar for useful comments that have considerably improved the manuscript. 

%%%%%%%%%%%%%%%%%%%%%%%%%%%%%%%%%%%%%%%%%%%%%%%%%%

\section*{DATA AVAILABILITY}

The data underlying this article will be shared on reasonable request to the corresponding author.



%\bibliographystyle{mnras}

\begin{thebibliography}{99}

%\bibitem[()]{}Abdo A. A. et al., 2013, ApJS, 208, 17

%\bibitem[()]{}Alpar, M. A., Cheng, A. F., Ruderman, M. A., \& Shaham, J. %1982, Nature,
%300, 728

\bibitem[()]{}Aly J. J., 1985, A\&A, 143, 19

\bibitem[()]{}Arons, J. 1993, ApJ, 408, 160

\bibitem[()]{}Archibald, A. M., Stairs, I. H., Ransom, S. M., et al. 2009, Sci, 324, 1411

%\bibitem[()]{}Archibald, A. M., Kaspi, V. M., Bogdanov, S., et al. 2010, ApJ, 722, 88

%\bibitem[()]{}Archibald, A. M., Kaspi, V. M., Hessels, J. W. T., et al. 2013, arXiv:1311.5161

%\bibitem[()]{}Archibald A. M. et al., 2015, ApJ, 807, 62

\bibitem[()]{}Bassa, C. G., Patruno, A., Hessels, J. W. T., et al. 2014, MNRAS, 441, 1825

%\bibitem[()]{}Bogdanov S., Archibald A. M., Hessels J. W. T., Kaspi V. M., Lorimer D.,McLaughlin M. A., Ransom S. M., \& Stairs I. H., 2011, ApJ, 742, 97

%\bibitem[()]{}Bogdanov, S., Esposito, P., Crawford, F., et al. 2014, ApJ, 781, 6

%\bibitem[()]{}Bogdanov, S., Archibald, A. M., Bassa, C., et al. 2015, ApJ, 806, 148

%\bibitem[()]{}Bogdanov, S., Deller, A. T., Miller-Jones, J. C. A., et al. 2018, ApJ, 856, 54 

\bibitem[()]{}Beri A., Jain C., Paul B., Raichur H., 2014, MNRAS, 439, 1940

\bibitem[()]{}Bildsten, L., et al. 1997, ApJS, 113, 367

\bibitem[()]{}Camero-Arranz, A., Finger, M. H., Ikhsanov, N. R., Wilson-Hodge, C. A., \& Beklen, E. 2010, ApJ, 708, 1500

\bibitem[()]{}Camero-Arranz, A., Finger, M. H., Wilson-Hodge, C. A., et al. 2012, ApJ,
754, 20

\bibitem[()]{}Chakrabarty, D., \& Roche, P., 1997, Apj, 489, 254

\bibitem[()]{}Chakrabarty, D., et al. 1997a, ApJ, 474, 414

\bibitem[()]{}Chakrabarty, D., et al. 1997b, ApJ, 481, L101

%\bibitem[D’Angelo\& Spruit( 2010)]{DS10}D'Angelo C. R., Spruit H. C., 2010, MNRAS, 406, 1208

%\bibitem[D’Angelo\& Spruit( 2011)]{DS11}D'Angelo C. R., Spruit H. C., 2011, MNRAS, 416, 893

%\bibitem[()]{}Crawford, F., Lyne, A. G., Stairs, I. H., et al. 2013, ApJ, 776, 20

\bibitem[D’Angelo\& Spruit( 2012)]{DS12}D'Angelo C. R., Spruit H. C., 2012, MNRAS, 420, 416

\bibitem[Davidson\& Ostriker( 1973)]{DO73}Davidson K., Ostriker J. P., 1973, ApJ, 179, 585

\bibitem[()]{}Deeter, J. E., Boynton, P. E., Lamb, F. K., \& Zylstra, G. 1989, ApJ, 336, 376

%\bibitem[()]{}Deller, A. T., Archibald, A. M., Brisken, W. F., et al. 2012, ApJL, 756, L25

%\bibitem[()]{}Done C., Gierli$\acute{n}$ski M., \& Kubota A., 2007, A\&AR, 15, 1

\bibitem[()]{}Ertan, \"{U}., 2017, MNRAS, 466, 175

\bibitem[()]{}Ertan, \"{U}., 2018, MNRAS, L12, 175

%\bibitem[()]{}Frank, J., King, A., \& Raine, D. 2002, Accretion Power in Astrophysics (3rd ed.; Cambridge: Cambridge Univ. Press)

\bibitem[()]{}Fromang S., \& Stone J. M., 2009, A\&A, 507, 19

%\bibitem[()]{}Gentile, P. A., Roberts, M. S. E., McLaughlin, M. A., et al. 2014, ApJ, 783, 69

\bibitem[()]{}Giacconi, R., Murray, S., Gursky, H., Kellogg, E., Schreier, E., \& Tananbaum,
H. 1972, ApJ, 178, 281

\bibitem[()]{}Ghosh, P., \& Lamb, F. K., 1979, ApJ, 234, 296

\bibitem[()]{}Gonz$\acute{\mathrm{a}}$lez-Gal$\acute{\mathrm{a}}$n, A., Kuulkers, E., Kretschmar, P., et al. 2012, A\&A, 537, A66

%\bibitem[()]{}Goodson A. P., Winglee R. M., \& Boehm K., 1997, ApJ, 489, 199 Negueruela et al. 2000, 

\bibitem[()]{}Hayashi M. R., Shibata K., \& Matsumoto R., 1996, ApJ, 468, L37

\bibitem[()]{}Hinkle, K. H., Fekel, F. C., Joyce, R. R., Wood, P. R., Smith, V. V., \& Lebzelter, T. 2006, ApJ, 641, 479

\bibitem[()]{}Illarionov A. F., Sunyaev R. A., 1975, A\&A, 39, 185

\bibitem[()]{}\.Inam, S. \c{C}.,  \c{S}ahiner, S.,\& Baykal, A., 2009, MNRAS, 395, 1015

\bibitem[()]{}Jaodand, A., Archibald, A. M., Hessels, J. W. T., et al. 2016, ApJ, 830, 122

\bibitem[()]{}Lamb, F. K., Pethick, C. J., \& Pines, D. 1973, ApJ, 184, 271 

%\bibitem[()]{}Li K. L., Kong A. K. H., Takata J., Cheng K. S., Tam P. H. T., \& Hui C. Y., Jin R., 2014, ApJ, 797, 111

%\bibitem[()]{}Linares, M. 2014, ApJ, 795, 72

\bibitem[()]{}Lovelace R. V. E., Romanova M. M., \& Bisnovatyi-Kogan G. S., 1995, MNRAS, 275, 244

\bibitem[()]{}Lovelace, R. V. E., Romanova, M. M., \& Bisnovatyi-Kogan, G. S., 1999, ApJ 514, 368

%\bibitem[()]{}Lyne, A.G., \& Graham-Smith, F. 2006 (ed.), in Pulsar Astronomy (3rd ed., Cambridge Astrophysics Series; Cambridge University Press), 264

\bibitem[()]{}Miller K. A., \& Stone J. M., 1997, ApJ, 489, 890

\bibitem[()]{}Orlandini, M., et al. 1998, ApJ, 500, L163

\bibitem[()]{}Ostriker, E. C., \& Shu, F. H. 1995, ApJ, 447, 813

\bibitem[()]{}Papitto, A., Ferrigno, C., Bozzo, E., et al. 2013a, Natur, 501, 517

\bibitem[()]{}Papitto A., de Martino D., Belloni T. M., Burgay M., Pelliz-
zoni A., Possenti A., \& Torres D. F., 2015, MNRAS, 449, L26

\bibitem[()]{}Petterson, J. A., Rothschild, R. E., \& Gruber, D. E. 1991, ApJ, 378, 696

%\bibitem[()]{}Patruno, A., \& D'Angelo, C. 2013, ApJ, 771, 94

%\bibitem[()]{}Pletsch H. J., \& Clark C. J., 2015, ApJ, 807, 18

%\bibitem[()]{}Radhakrishnan, V., \& Srinivasan, G. 1982, CSci, 51, 1096

%\bibitem[Rappaport et al.(2004)]{Rappaport04} Rappaport, S. A., Fregeau, J. M., \& Spruit, H. 2004, ApJ, 606, 436

%\bibitem[()]{}Rothschild, R., Markowitz, A., Hemphill, P., et al. 2013, ApJ, 770, 19

%\bibitem[()]{}Rutledge, R. E., Bildsten, L., Brown, E. F., et al. 2007, ApJ, 658

%\bibitem[()]{}Ray, P. S., Belfiore, A. M., Saz Parkinson, P., et al. 2015, in American
Astronomical Society Meeting Abstracts, 223, 140.07

%\bibitem[()]{}Romanova, M. M., Ustyugova, G. V.,  Koldoba, A. V., \& Lovelace, R. V. E., 2005, ApJ, 635, L165

%\bibitem[()]{}Romani, R. W., \& Shaw, M. S. 2011, ApJL, 743, L26

\bibitem[Shakura \& Sunyaev(1973)]{SS73}Shakura, N. I., \& Sunyaev, R. A. 1973, A\&A, 24, 337

%\bibitem[()]{}Spruit H. C., \& Taam R. E., 1993, ApJ, 402, 593

%\bibitem[()]{}Stappers, B. W., Archibald, A. M., Hessels, J. W. T., et al. 2014, ApJ, 790, 39

%\bibitem[()]{}Sunyaev, R. A., \& Shakura, N. I. 1977, PAZh,
3, 262

\bibitem[()]{}Ustyugova, G. V., Koldoba, A. V., Romanova, M. M., \& Lovelace, R. V. E., 2006, ApJ, 646, 304

\bibitem[()]{}Uzdensky, D. A., K\"{o}nigl A., \& Litwin C., 2002, ApJ, 565, 1191

\bibitem[()]{}Uzdensky, D. A., 2004, Ap\&SS, 292, 573

%\bibitem[()]{}Wang, Y.-M. 1987, A\&A, 183, 257


\end{thebibliography}


%%%%%%%%%%%%%%%%%%%%%%%%%%%%%%%%%%%%%%%%%%%%%%%%%%

%%%%%%%%%%%%%%%%% APPENDICES %%%%%%%%%%%%%%%%%%%%%

%\appendix



%%%%%%%%%%%%%%%%%%%%%%%%%%%%%%%%%%%%%%%%%%%%%%%%%%

% Don't change these lines
\bsp	% typesetting comment
\label{lastpage}
\end{document}